\documentclass{pasj01}
\draft 

\begin{document}
\Received{}
\Accepted{}

\title{A systematic comparison of ionization temperatures between ionizing and recombining plasmas in supernova remnants}

\author{
Shigeo \textsc{Yamauchi}\altaffilmark{1},
Masayoshi \textsc{Nobukawa}\altaffilmark{2},
and 
Katsuji \textsc{Koyama}\altaffilmark{3}
}
\altaffiltext{1}{Faculty of Science, Nara Women's University, Kitauoyanishimachi, Nara 630-8506, Japan}
\altaffiltext{2}{Faculty of Education, Nara University of Education, Takabatake-cho, Nara 630-8528, Japan}
\altaffiltext{3}{Department of Physics, Graduate School of Science, Kyoto University, \\
Kitashirakawa-oiwake-cho, Sakyo-ku, Kyoto 606-8502, Japan}

\email{yamauchi@cc.nara-wu.ac.jp}


\maketitle

\begin{abstract}

The temperatures of the plasma in the supernova remnants (SNRs) are initially very low just after the shock heating. 
The electron temperature ($kT_{\rm e}$) increases quickly by Coulomb interaction, and 
then the energetic electrons gradually ionize atoms to increase the ionization temperature ($kT_{\rm i}$). 
The observational fact is that most of the young and middle-to-old aged SNRs have lower $kT_{\rm i}$ than $kT_{\rm e}$ after the shock heating. 
The temperature evolution in the shell-like SNRs has been explained by this ionizing plasma (IP) scenario. 
On the other hand, in the last decade, a significant fraction of the mixed morphology SNRs was found to exhibit a recombining plasma (RP) 
with higher  $kT_{\rm i}$ than $kT_{\rm e}$. 
The origin and the evolution mechanism of the RP SNRs have been puzzling.  
To address this puzzle, this paper presents the $kT_{\rm e}$ and  $kT_{\rm i}$ profiles using the observed results by follow-up Suzaku observations, 
and then proposes a new scenario for the temperature and morphology evolutions in the IP and RP SNRs.

\end{abstract}

\section{Introduction} 
 
The X-ray emitting thermal plasmas are found in the shell-like supernova remnants (SNRs) and mixed morphology (MM) SNRs \citep{Rho1998}. %
The three important physical parameters that characterize the spectrum are
the electron temperature  ($kT_{\rm e}$), ion (proton) temperature ($kT_{\rm p}$), and ionization temperature ($kT_{\rm i}$). 
The evolution of thermal plasma of the shell-like SNRs starts from shock heating by the blast wave (called the {\it Adiabatic~phase}) 
following the free expansion (called the {\it Free~Expansion~phase}). 
At the start epoch of the {\it Adiabatic~phase}, the temperatures of $kT_{\rm e}$ and $kT_{\rm i}$ are very low. $kT_{\rm p}$ is also low, 
but is higher than $kT_{\rm e}$  because of higher energy than electrons with larger mass of ion (proton). %
In the {\it Adiabatic~phase}, the protons transfer their energy to the electrons, 
and increase $kT_{\rm e}$ in a short time scale. 
Then the electrons gradually ionize atoms to increase $kT_{\rm i}$. Thus, most of the young shell-like SNRs have plasma of 
$kT_{\rm e} > kT_{\rm i}$, called ionizing plasma (IP) (e.g., \cite{Masai1994,Truelove1999}). 
After a long time of evolution, the middle-to-old aged shell-like SNRs are still in $kT_{\rm e} > kT_{\rm i}$ (IP), 
or become $kT_{\rm e} \sim kT_{\rm i}$ (collisional ionization equilibrium, CIE) at the end of the {\it Adiabatic~phase}, 
just before the {\it Radiative~Cooling~phase}. 

This evolution scenario of  the IP in the shell-like SNR is well established in the theory (model) and observation (e.g., \cite{Masai1994,Truelove1999, Ts86}). 
In the last decade, $kT_{\rm e} < kT_{\rm i}$ plasma (recombining plasma, RP) was found in a significant fraction of MM SNRs 
(e.g., \cite{Hirayama2019}, and references therein). 
One of the conventional scenarios is that cold clouds are responsible for both the MM 
and RP; the evaporation of cold clouds makes extra thermal X-rays in the inner region 
of the SNR, transforming the shell structure to the MM \citep{White1991}, and the same
cold clouds decrease $kT_{\rm e}$ by thermal conduction, converting the IP into RP. 

\citet{Itoh1989} proposed the rarefaction model, a unique model to explain the full evolution from the IP to RP SNRs.
The adiabatic expansion to low density space causes rapid cooling of $kT_{\rm e}$ and makes RP.
In the rarefaction model, the dense layer of circumstellar matter (CSM) causes reverse-shock, 
makes a higher post-shock temperature than a case without a dense CSM, 
and produces low-energy cosmic ray protons (LECRp) by the same process of the diffusive shock acceleration in the blast wave \citep{Sh13}. 
The rarefaction model, however, may not work for all the RP SNR evolution; in actual SNRs, there may be many competing processes 
for the conversion of the IP into the RP. 
The cloud evaporation would be one example which is not considered in the rarefaction model. 
It may decrease $kT_{\rm i}$, because of an enhanced recombination by cool electrons in the dense CSM. 
Thus, although the rarefaction is a unique model at present that can successfully applied for the evolution to RP in some of the IP SNRs, 
it may not for all the RP SNRs. 

To evaluate which processes are actually operating in the evolution of  RP SNRs, we propose a new scenario, 
which is free from theory and/or numerical calculation. 
We make the $kT_{\rm e}$ and $kT_{\rm i}$ profiles, which reflect the balance between the two competing processes, the increase of $kT_{\rm i}$ 
(enhanced ionization by LECRp) and the decrease of $kT_{\rm i}$ (enhanced recombination because of cool electron from evaporated clouds). 
As for the $kT_{\rm e}$, there are complementary processes for electron cooling: adiabatic expansion, heat conduction from cold clouds, 
and energy loss by ionization. 
The observation-based temperature profiles lead us to a new scenario for the evolution of the RP SNRs. 

This paper organizes as follows. 
In the next section (section 2), high quality spectra from 5 IP and 4 RP SNRs are taken from the archives of Suzaku \citep{Mitsuda2007}. 
Section 3 gives the method and formalism of the spectral analysis for these SNRs, 
and then presents the fitting results of $kT_{\rm i}$ and $kT_{\rm e}$ for IP and RP SNRs. 
These results are complied, and are used to construct the transition of RP from IP in section 4. 
Section 5 is assigned for the discussion in comparison between the conventional model and the new scenario, 
and follows to the future prospect for the advanced picture of IP and RP evolutions.

\section{Selection of Archive Data} 

For comparison of IP and RP, we utilized the Suzaku archive data from the X-ray Imaging Spectrometer (XIS: \cite{Ko07}) 
placed at the focal planes of the thin foil X-ray Telescopes (XRT: \cite{Se07}). 
We have selected the most reliable 5 IP and 4 RP SNRs, which are bright and/or have enough observational time, 
so that the best-fit physical parameters have small errors to precisely determine the SNR parameters. 
Among these, young IP SNRs  are Cas A, Tycho, and SN 1006, middle-to-old aged IP SNRs are CTB 109 and Cygnus Loop, young RP SNR is W49B, and  middle-to-old aged RP SNRs are IC443, W28, and G359.1$-$0.5. %
The observational logs of these SNRs are listed in table 1.

\begin{table*} 
\caption{The observation logs of the selected SNRs.}
\label{tab:first}
\begin{center}
 \begin{tabular}{cccc}
\hline 
Position & Observation ID & Start -- End & Exposure time (ks)\\ \hline
\multicolumn{4}{c}{IP SNRs}\\ \hline

\multicolumn{4}{c}{Cas A} \\
	&100016010	&2005-09-01 05:58:03 -- 2005-09-01 18:20:14& 28.0 \\ 

\multicolumn{4}{c}{Tycho} \\
	&500024010&2006-06-27 10:32:29 -- 2006-06-29 15:40:24& 101.1 \\ 

\multicolumn{4}{c}{SN 1006} \\
SE& 500016010 & 2006-01-30 09:01:20 -- 2006-01-31 11:42:14 & 51.6\\

\multicolumn{4}{c}{CTB 109} \\
NE &   506039010 &  2011-12-15 01:57:25 -- 2011-12-15 18:03:11  & 30.4 \\
SE &   506040010 &  2011-12-15 18:03:52 -- 2011-12-16 09:37:06  & 30.4 \\

\multicolumn{4}{c}{Cygnus Loop} \\
NE1& 500020010 & 2005-11-23 17:39:01 -- 2005-11-24 04:55:24 & 20.4\\
NE2& 500021010 & 2005-11-24 04:56:05 -- 2005-11-24 16:14:24 & 21.4 \\
NE3& 500022010 & 2005-11-29 17:47:47 -- 2005-11-30 05:39:09 & 21.7\\
NE4& 500023010 & 2005-11-30 05:41:02 -- 2005-11-30 18:23:14 & 25.3\\

\hline 
\multicolumn{4}{c}{RP SNRs}\\ \hline
\multicolumn{4}{c}{W49B} \\
&	503084010&2009-03-29 02:33:12 -- 2009-03-30 11:15:18 & 52.2\\
&	503085010&2009-03-31 12:43:35 -- 2009-04-02 01:28:20 & 61.4\\
\multicolumn{4}{c}{IC443} \\
NE &	501006010&2007-03-06 10:40:19 -- 2007-03-07 12:22:14&42.0 	\\
NE &	507015010&2012-09-27 05:29:48 -- 2012-09-29 18:40:22 &101.8	\\
NE &	507015020&2013-03-27 04:15:06 -- 2013-03-28 16:00:19& 59.3	\\
NE &	507015030&2013-03-31 11:44:34 -- 2013-04-03 21:12:21& 131.2	\\
NE &	507015040&2013-04-06 05:21:49 -- 2013-04-08 02:00:21 & 75.6	\\
\multicolumn{4}{c}{W28} \\
Center &505005010&2010-04-03 07:23:22 -- 2010-04-04 23:48:14&73.0\\
E &505006010&2011-02-25 10:54:11 -- 2011-02-28 04:08:07& 100.0\\
\multicolumn{4}{c}{G359.1$-$0.5} \\
W& 	502016010&	2008-03-02 18:08:00 -- 2008-03-04 17:40:19&	70.5\\
S&	502017010&	2008-03-06 13:26:36 -- 2008-03-08 16:00:24& 	72.5\\
N& 	503012010&	2008-09-14 19:35:07 -- 2008-09-16 00:50:14&	57.7\\
\hline
\end{tabular}
\end{center}
\end{table*}

\section{Analysis} 
\subsection{Method of spectrum fit} 

To derive a unified picture for the evolution of RP and IP SNRs, we uniformly fitted the X-ray spectra of the selected nine SNRs ({\it unified~fit}).
The X-ray spectra of the SNRs (source spectra) were obtained after subtracting the non X-ray background (NXB, \cite{Tawa2008}) and off source sky background spectra (XB). 
The XB is either the Galactic off-plane spectra or the Galactic ridge spectra made by the method of 
\citet{Masui2009} (NXB) or \citet{Uchiyama2013} (XB) using  the off-source sky region in the source fields listed in table 1.

The SNR spectra (source spectra) were fitted with a model having  $kT_{\rm i}$ for each element, 
$kT_{\rm i}(z)$, in the {\it VVRNEI} code of the {\it XSPEC} package (multi-{\it VVRNEI} model).  
The multi-{\it VVRNEI} is $z$ dependent code, where the relevant elements were grouped to Mg--Ar (the group A). 
The other elements were grouped in the order of the atomic number ($z$), He--Ne is the group B and Ca--Zn is the group C.  

The fitting model using  multi-{\it VVRNEI}  code is
\begin{equation}
\sum_{\rm z=H}^{\rm z=Ni}{\it VVRNEI} [kT_{\rm e}, kT_{\rm i}(z), n_{\rm e}t, {\it redshift}],
\end{equation} 
where $n_{\rm e}$ and $t$ are the plasma density (cm$^{-3}$) and the evolution time (s), respectively. 
The other free parameter {\it redshift} is used for the fine-tuning of possible time-dependent energy scale (due to the calibration uncertainty) 
at the emission lines of the relevant elements. 

In order to discuss the process of the evolution from IP to RP as essentially and clearly as possible, the spectral structure of Mg--Ar (the group A) 
in the energy band of 1.3--4 keV was used because the essential difference between IP and RP was found near the K shell lines of these elements. 
In the fitting, the free parameters in the group A were $kT_{\rm e}$, $kT_{\rm i}(z)$, $n_{\rm e}t$, and abundances. 
In the group B, those of $kT_{\rm e}$, $kT_{\rm i}$, and $n_{\rm e}t$ were linked to those of Mg. 
The group C was same as the group B, but the free parameters were linked to those of Ar. 
We assumed all the abundances in the groups B and C to be $\sim$1 solar. %

Although the energy resolution in the early observation was not degraded by particle background, 
those of the late observations were significantly degraded. 
The line broadening due to these time-dependent variations of the energy resolution and due to the spectrum sum are $\sim$30 eV (FWHM). 
To compensate these line broadenings, we applied the {\it gsmooth} code in the {\it XSPEC} package. 

The spectra of many data sets (table 1) were merged into the final source spectra. 
The line broadening due to this merging effect can also be compensated by  the {\it gsmooth} code.
The spectrum of G359.1$-$0.5 with the {\it unified~fit} is given in figure 1 as a typical example of the selected RP SNRs. 
We see clear data excess at the energy of He$\alpha$ and Ly$\alpha$ and radiative recombination continuum (RRC) of Si and S above the CIE model, 
as is shown in figure 1b. 
Here, we call these excess as the Radiative Recombination Structure (RRS). 
The RRS was found in the residual of  the CIE fit with $\chi^2$/d.o.f. of 629/167 (figure 1b), 
and is drastically disappeared in the RP model fit with $\chi^2$/d.o.f. of 206/165 (figure 1c). 
Therefore, the RRSs of Si and S can be used  to judge whether the SNRs are RP or IP (or CIE). 
The sampled MM RP SNRs (W49B, IC443, W28, and G359.1$-$0.5) exhibited the most clear RRSs among the known MM RP SNRs. 

\begin{figure}[tb] 
\begin{center}
\includegraphics[clip, width=8cm]{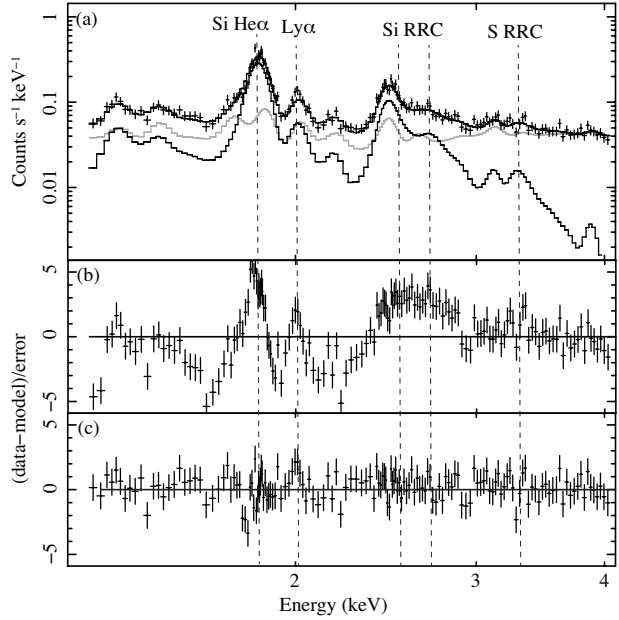} 
\end{center}
\caption {(a) The spectrum of an RP SNR, G359.1$-$0.5 (sum of the XIS\,0 and 3 data, \cite{Ko07}). 
The black and gray curves are the RP model for G359.1$-$0.5 and the background model, respectively. 
The dashed lines indicate the energies of Si He$\alpha$ and Ly$\alpha$ lines and the edge energies of Si and S RRCs. 
(b) Residuals of the spectrum for the CIE model (see text). (c) Same as (b), but for the RP model.} 
\label{IP-RP_Curve}   
\end{figure}

\subsection{The multi-{\it VVRNEI} fit} 

Using the multi-{\it VVRNEI} code, we carried out two model fittings, $n_{\rm e}t$ = 0 fit and  free-$n_{\rm e}t$ fit, for the IP SNRs. 
The $n_{\rm e}t$ values in the free-$n_{\rm e}t$ fit correspond to the duration of the spectral evolutions from the starting epoch to the observed epoch. 
The best-fit $kT_{\rm e}$ and $kT_{\rm i}(z)$ of Mg, Si, S, and Ar in the $n_{\rm e}t$ = 0 fit (observed epoch) 
and $n_{\rm e}t$ in the free-$n_{\rm e}t$ fit for each IP SNRs are listed in table 2. 
We also fitted the RP SNR spectra with fixed $n_{\rm e}t$ = 0 and free-$n_{\rm e}t$. 
The $n_e t$ values obtained from the free-$ n_{\rm e}t$ fit for the RP SNRs do not correspond  the SNR ages, but correspond the elapsed time after the transition to the RP phase. 
The best-fit parameters are listed in table 3. 

\begin{table*} 
\caption{The best-fit parameters for IP SNRs.}
\label{tab:first}
\begin{center}
\begin{tabular}{cccccc}
\hline
				&Cas A 			&Tycho  				&SN 1006 			&CTB 109 			&Cygnus Loop\\ \hline
\multicolumn{6}{c}{Observed epoch ($n_{\rm e}t$ = 0 cm$^{-3}$s)} \\ \hline
$kT_{\rm e}^*$		&$3.09\pm{0.04}$ 	&$1.76\pm{0.02}$ 		&$0.98\pm{0.04}$		&$0.57\pm{0.03}$ 	 	&$0.30\pm{0.01}$\\
$kT_{\rm i}$(Mg)$^*$ &$0.59\pm{0.01}$	&$0.50\pm{0.02}$		&$0.09\pm{0.01}$		&$0.49\pm{0.02}$		&$0.30\pm{0.05}$\\
$kT_{\rm i}$(Si)$^*$ &$0.78\pm{0.01}$	&$0.56\pm{0.01}$ 		&$0.13\pm{0.01}$		&$0.57\pm{0.05}$		&$0.29\pm{0.05}$ \\		
$kT_{\rm i}$(S)$^*$   &$0.99\pm{0.01}$	&$0.68\pm{0.02}$		&$0.18\pm{0.01}$		&$^{\dag}$			&$^{\dag}$\\
$kT_{\rm i}$(Ar)$^*$  &$1.47\pm{0.01}$	&$0.84\pm{0.01}$ 		&$0.25\pm{0.03}$		&$^{\dag}$			&$^{\dag}$\\
$\chi^2$/d.o.f.    	&1721/744 (2.31)	&2577/720 (3.58)		&447/373 (1.20)		&343/265 (1.29)		&269/269 (1.00)\\ \hline
\multicolumn{6}{c}{$n_{\rm e}t$=free} \\ \hline
$n_{\rm e}t$ (cm$^{-3}$ s) &
$(5.2\pm0.1)\times10^{10}$ & $(2.9\pm0.1)\times10^{10}$ & $(6.3\pm0.7)\times10^{9}$  & $(6.3\pm0.8)\times10^{11}$	& $(3.0\pm1.7)\times10^{12}$ \\ 
\hline
\end{tabular}
\end{center}
Assuming the plasma density, $n_{\rm e}$ (cm$^{-3}$), is 1 cm$^{-3}$, $t$ (s) from the best-fit $n_{\rm e}t$ value in the $n_{\rm e}t$ = free fitting 
is the age of the IP SNR. \\
$^{\ast}$ Units are keV. \\
$^{\dag}$ not determined due to poor data in the high $z$ elements.\\
\end{table*}

\begin{table*} 
\caption{The best-fit parameters for RP SNRs.}
\label{tab:first}
\begin{center}
 \begin{tabular}{ccccc}
 \hline
					&W49B  				&IC443				&W28 			&G359.1$-$0.5 \\ \hline
\multicolumn{5}{c}{Observed epoch  ($n_{\rm e}t$ = 0 cm$^{-3}$s)} \\ \hline
$kT_{\rm e}^*$			&$0.75\pm{0.01}$ 		&$0.55\pm{0.01}$ 		&$0.33\pm0.05$ 	&$0.16\pm0.02$		\\
$kT_{\rm i}$(Mg)$^*$	&$0.75\pm{0.01}$		&$0.69\pm{0.01}$  		&$0.63\pm0.03$	&$0.53\pm0.04$		\\
$kT_{\rm i}$(Si)$^*$		&$1.21\pm{0.01}$		&$1.10\pm{0.01}$  		&$1.03\pm0.06$	&$0.77\pm0.02$		\\		
$kT_{\rm i}$(S)$^*$		&$1.60\pm{0.01}$		&$1.33\pm{0.01}$ 		&$1.20\pm0.07$	&$0.85\pm0.21$		\\
$kT_{\rm i}$(Ar)$^*$		&$2.03\pm{0.03}$		&$1.00\pm{0.01}$ 		&$1.14\pm0.22$	&$0.83\pm0.52$		\\
\hline
$\chi^2$/d.o.f.			&558/344 (1.62)		&1148/583 (1.97) 		&365/335 (1.09)	&381/363 (1.05) \\
\hline
\multicolumn{5}{c}{$n_{\rm e}t$=free} \\ \hline
$n_{\rm e}t$ (cm$^{-3}$ s) &
$(2.8\pm0.2)\times10^{11}$&$(2.7\pm0.1)\times10^{11}$&$<1.0\times10^{11}$&$1.5^{+5.0}_{-1.4} \times 10^{11}$\\
\hline
\end{tabular}
\end{center}
Assuming the plasma density, $n_{\rm e}$ (cm$^{-3}$), is 1 cm$^{-3}$, $t$ (s) from the best-fit $n_{\rm e}t$ value in the $n_{\rm e}t$ = free fitting 
is the duration time of the RP phase of the RP SNR. \\
$^*$ Units are keV.
\end{table*}

\section{Results and unified picture of IP and RP SNRs} 
 
\subsection{The $kT_{\rm i}$ distribution with $z$ in the IP and RP SNRs} 

The $kT_{\rm i}$ distribution of each SNRs as a function of $z$ is plotted in figure 2.  We find notable facts as below. 

\medskip
(1) The $kT_{\rm i}$ values in both the IP and RP SNRs are smaller in the smaller $z$, and have similar shape with each other. 

(2) The $kT_{\rm i}$ values of RP SNRs are larger than those of IP SNRs. 

\medskip

\citet{Sawada2012} compared the element dependent $kT_{\rm i}~(z)$-profile for CIE and IP process in the spectrum of RP SNR W28. 
The element dependent $kT_{\rm i}~(z)$-profile in the RP SNR W28 is consistent with our results of (1).
The facts (1) and (2) naturally lead to the idea that the RP SNR comes after the well established phase in the shell-like IP SNR. 
In this idea, the RP SNRs phase smoothly follows after an IP process (IP SNR). 

\begin{figure}[tb] 
\begin{center}
\includegraphics[width=8cm]{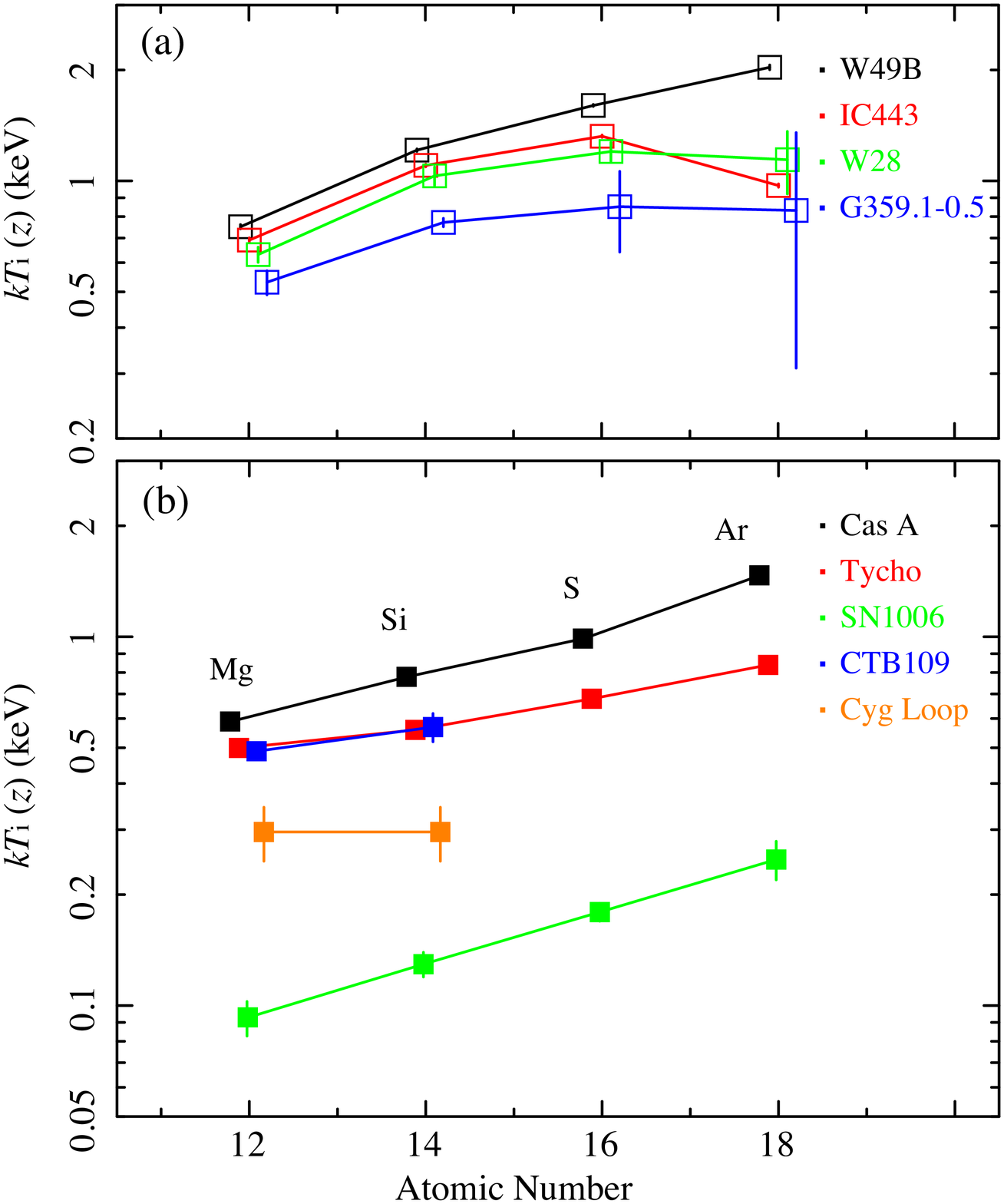} 
\end{center}
\caption {The $kT_{\rm i}(z)$ distribution of selected (a) RP and (b) IP SNRs as a function of $z$.
} 
\end{figure}

\subsection{The $kT_{\rm e}$ distribution in the IP SNRs} 

The X-ray spectra of the SNR plasma is mainly  composed of line and continuum emissions. The former, mainly He$\alpha$ and Ly$\alpha$, is determined by $kT_{\rm i}$, while the latter is determined by $kT_{\rm e}$. 
Thus, $kT_{\rm e}$ represents the main component in the hot plasma, almost independent of $kT_{\rm i}$ (insensitive to IP or RP).  
The evolution curve of the similarity solution \citep{Sedov1959} predicts that the $kT_{\rm e}$ is a power-law function with the index of $6/5$ 
as a function of age ($t$). 
However, the real $kT_{\rm e}$ evolution curve may be significantly modified by the thermal conduction from cold cloud, energy transfer from proton 
($kT_{\rm p}>kT_{\rm e}$ in young SNRs, \cite{Rakowski2003}), energy dependent competing process (ionization and recombination), or else. 
Details of these processes are complicated, hence not well predictable. 
Furthermore, these processes are coupled with each other, and hence theoretical prediction of the $kT_{\rm e}$ evolution curve is not available at present. 
We therefore make the $kT_{\rm e}$ evolution curve of IP SNRs at the best-fit position of the $n_{\rm e}t$ = 0 fit (black solid line in figure 3). 
For young IP SNRs (Cas A, Tycho, and SN 1006), well established ages were used, while those for old aged SNRs 
(CTB 109 and Cygnus Loop) were estimated using the best-fit value of $t=n_{\rm e}t/n_{\rm e}$. 
Since the evolution of $kT_{\rm e}$ in the IP SNR starts from the initial  epoch of the {\it Adiabatic phase}, $t$ is regraded approximately to be the SNR age, if $n_{\rm e}$ is in typical value of 1 cm$^{-3}$.
This evolution curve shows a monotonous decrease with the age $t$, qualitatively similar to the Sedov solution.
As we noted before, the $kT_{\rm e}$ evolution curve of the RP SNR should be nearly similar to that of the IP SNRs. 
The youngest RP W49B has $kT_{\rm e}$$\sim$0.8 keV in the $n_{\rm e}t$=0 fit at the epoch of observing time (table 3).  

As for $kT_{\rm e}$ of W49B, \citet{Oz09} and \citet{Ya18} reported to be $\sim$1.1--1.8 keV, nearly 1.3--2.3 times of $\sim$0.8 keV.
However, these values were determined in the hard energy band of 5--12 keV \citep{Oz09} or 3--20 keV \citep{Ya18}. 
On the other hand, this paper used softer band of 1.3--4 keV. 
Furthermore, the $kT_{\rm e}$ values of \citet{Oz09} and \citet{Ya18} are those at the start epoch of the RP phase ($n_{\rm e}t$-free fit), 
while this paper is at the observed epoch ($n_{\rm e}t$=0 fit). 
Thus, the larger $kT_{\rm e}$ of \citet{Oz09} and \citet{Ya18} than this paper may be reasonable.  
The $kT_{\rm e}$ value of W49B corresponds to the $kT_{\rm e}$ of the IP SNR between SN 1006 and CTB 109. 
Therefore, the best-fit $kT_{\rm e}$ of W49B (RP) is placed on the $kT_{\rm e}$ evolution curve of the IP SNRs 
at the position near SN 1006 and CTB 109.  
Likewise, we placed the $kT_{\rm e}$ values of middle-to-old aged RP SNRs (IC443, W28, 
and G359.1$-$0.5) near to the positions of the middle-to-old aged IP SNRs (CTB 109 and Cygnus loop). 

In the young IP SNRs, $kT_{\rm i}$ is  smaller than $kT_{\rm e}$, while in the old SNRs (CTB 109 and Cygnus Loop), $kT_{\rm i}$ becomes almost equal to $kT_{\rm e}$ (figure 3). 
The $kT_{\rm e}$ is peaked at Cas A (age$\sim$300 year), and monotonously decreases as increasing age. 
This indicates that the {\it Adiabatic phase} starts near at 300 years after the {\it Free Expansion phase} \citep{Sedov1959}.
In the evolution of  middle-to-old aged IP SNRs, the epoch when the $kT_{\rm i}$ (Si) value becomes equal to that of $kT_{\rm e}$ is at the age of CTB 109. 
This indicates that the ionization time scale is $\sim2\times10^4$ year, which is significantly smaller than the recombination time scale. 
The evolution continues keeping the balance of recombination rate$\simeq$ ionization rate.  

\subsection{Difference of the $kT_{\rm i}$ (Si)-profile between the IP and RP SNRs} 

The $kT_{\rm i}$ (Si)-profile of IP SNRs are given in figure 3 with the filled square, 
while the red solid line is to connect each data (the $kT_{\rm i}$ (Si)-profile).  
The open squares in figure 3 are $kT_{\rm i}$ (Si)-profile of the RP SNRs, 
while the blue line is to connect each data. 
The $kT_{\rm i}$ (Si)-profile must be different between IP and RP SNRs, 
because  the line fluxes and the RRS of key element Si are due to the local two-body process in ionization and/or recombination. 
As we mentioned in section 4.2, Cas A, Tycho, and SN1006 are historical SNRs, and hence the ages are either well predicted (Cas A) or 
actually observed date of the SNRs (Tycho and SN1006). 
For the other two, we estimated the ages using the best-fit $n_{\rm e}t$ values. 
On the other hand, we cannot find any reports for the age or elapsed time in RP phase. 
Thus, interpolating the $kT_{\rm e}$ values between IP and RP SNRs, we determined the horizontal positions for RP SNRs. %

The young RP SNR, W49B, would have changed to the RP phase due to the extra ionization. At the epoch of this phase change, $kT_{\rm i}$ (Si) ($\sim$1.2 keV) is higher  than $kT_{\rm e}$ ($\sim$0.8 keV). 
No $kT_{\rm i}$ data of RP SNRs that are younger than  $\sim10^3$ year old 
are found among our selected samples\footnote{The age estimation in this paper is based on the best-fit $n_{\rm e}t$.  As we noted, this age estimation is not reliable for the RP SNR, and hence should be regarded conservatively.}.
In the middle-to-old aged SNRs (IC443, W28, and G359.1$-$0.5), 
the $kT_{\rm i}$~(Si) profile is systematically higher than those of the middle-to-old aged IP SNRs (CTB 109 and Cygnus Loop).

\begin{figure}[tb] 
\begin{center} 
\includegraphics[width=8cm]{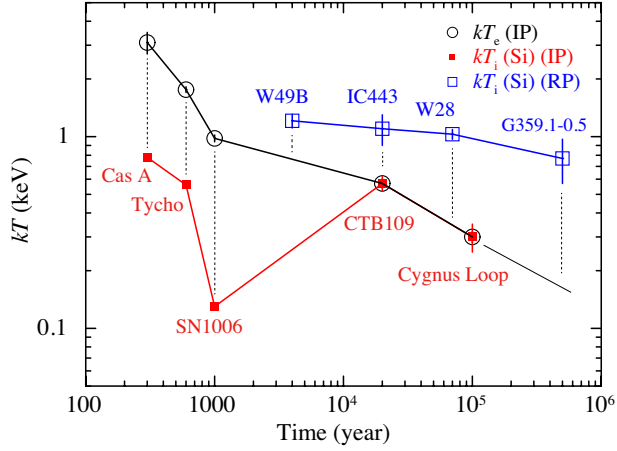} 
\end{center}
\caption {The empirical evolution curve for the selected samples of IP and RP.
The open and filled symbols are $kT_{\rm e}$ and  $kT_{\rm i}$ (Si). 
The evolution curve of $kT_{\rm e}$ for  IP in the adiabatic expansion phase and that of $kT_{\rm i}$ are shown by the black and red lines, respectively. 
The evolution curve of $kT_{\rm i}$ for RP is shown by the blue line.
The dashed lines are to guide eye for the evolution of  $kT_{\rm e}$ of IP and RP as a function of $n_{\rm e}t$. 
The values of horizontal axis are the ages of SNR (the epoch is starts of  the Adiabatic phase),
where the ages of  $>10^4$ year (CTB 109 and Cygnus Loop) are estimated age from  the best-fit $n_{\rm e}t$ for $n_{\rm e}$=1 cm$^{-3}$.}
\label{figure3 }  
\end{figure}

\section {Discussion and future prospect}  

Using the observed data and the results of the multi-{\it VVRNEI} model  fit, we made three evolution curves of 
$kT_{\rm e}$ (1), and $kT_{\rm i}$ (Si) of RP (2) and IP (3). 
A new idea of this scheme is that we can determine $kT_{\rm i}$ and $kT_{\rm e}$ at the observed epoch 
with the {\it VVRNEI} code of $n_{\rm e}t$ = 0 fit. 
We note that conventional {\it VVRNEI} analysis predicts no $kT_{\rm i}$ value at the observation epoch. 

In the $kT_{\rm e}$  profile (1), a critical assumption is that the $kT_{\rm e}$ curves of IP and RP are similar. 
This assumption is based on the fact that $kT_{\rm e}$ is determined by the global, dynamical evolution of the thermal plasma. 
One problem is that the age estimation (using the $n_{\rm e}t$ value in the free $n_{\rm e}t$ fit) of the middle-to-old aged SNRs 
has significant ambiguities and/or statistical errors.
For example, the best-fit $n_{\rm e}t$ values of CTB 109 and Cygnus Loop give 
their ages of $(2.0 \pm 0.3)\times10^4$ and  $(9.5\pm5.4)\times10^4$ year, respectively. 
Then within these errors, the important conclusion that $kT_{\rm e}$ has a monotonous decrease with slow rate 
as a function of SNR age $t$ (figure 3) is not changed. 

In the  $kT_{\rm i}$ (Si) profile of the RP SNRs (2), the $kT_{\rm i}$ value of young SNR ($< 10^4$ year) W49B is plotted as $\sim1$ keV 
at the position of the IP SNR-line of $kT_{\rm e}\sim0.8$ keV, 
while that in the old SNR ($ > 10^5$ year) G359.1$-$0.1 is plotted as $\sim0.7$ keV at the IP SNR-line of $kT_{\rm e}\sim0.2$ keV. 
Then, the $kT_{\rm i}$ (Si) of the RP SNRs shows very slow deceases as increasing $n_{\rm e}t$.

In the IP SNRs (3), unlike that in RP SNRs, the $kT_{\rm i}$ (Si) has a local minimum at the epoch of $t\sim10^3$ years (at the position of SN 1006). 
The lower ionization temperature in SN 1006 than the expectation from the overall trend would be due to the difference in the density. SN1006 is known to have a low density environment (e.g., \cite{Du02}) and therefore the thermal evolution of its shock-heated ejecta is slower than the others (e.g., \cite{Ya08}). 

In the old age of low $kT_{\rm e}$ region, the recombination rate becomes nearly equal to the ionization rate, 
hence $kT_{\rm i}$ (Si) saturates near at  the $kT_{\rm e}$ values (after the position of CTB 109). 
The observational fact that $kT_{\rm i}$ values of the RP SNRs are much higher than those of any IP SNRs implies that 
RP originated from some extra ionization rather than cooling of $kT_{\rm e}$ as suggested in the conduction and rarefaction models. 
Based on these 3 observational profiles, we propose a new scheme of the unified picture for the IP and RP evolution. 
In this scheme, an essential  player is  small cloud-lets. 
The small cloud-lets would produce LECRp by the diffusive shock acceleration in the hot plasma.  
LECRp is the most likely source to give additional ionization and then plasma changes from IP to RP (phase transition in ionization state), 
when the $kT_{\rm e}$ value of RP gradually becomes equal to that of IP in the full evolution history.
The cloud-lets also make diffuse hot plasma in the interior of SNR surrounded by the shell.
It causes another phase transition from the  shell-like to the MM SNR (phase transition in morphology).

The profiles of (1), (2), and (3) were made by the limited samples of SNR data with high quality spectra. 
This is a weak point in our scenario (biased picture). 
To make unbiased picture, we encourage to observe a greater numbers of IP and RP SNRs in high energy resolution  and statistics. %
The next Japanese-led mission, X-Ray Imaging and Spectroscopy Mission (XRISM), has 
high capability of X-ray spectroscopy using a micro-calorimeter, which would be suitable for such observations.
This can be powerful for the study of line flux and width in the energy band of the RRS, which would distinguish the SNRs whether it is IP or RP.  
It also provides key information for the  study of the origin of RP and transition mechanism of the IP to RP. 
Then the reliability of our unified scenario for the temperature and morphology evolution in both the IP and RP SNRs 
should become higher than that of our present scenario.


\begin{thebibliography}{}
\bibitem[Dubner et al. (2002)] {Du02} 
   Dubner, G. M., Giacani, E. B., Goss, W. M., Green, A. J., \& Nyman, L.-\AA.  2002,  A\&A, 387, 1047
\bibitem[Hirayama et al. (2019)]{Hirayama2019}  
   Hirayama, A., Yamauchi, S., Nobukawa, K. K., Nobukawa, M., \& Koyama, K. 2019, PASJ, 71, 37
\bibitem[Itoh \& Masai(1989)] {Itoh1989} Itoh, H., \& Masai, K. 1989, MNRAS, 236, 8851
\bibitem[Koyama et al.(2007)]{Ko07} 
Koyama, K., et al.\ 2007, \pasj, 59, S23
\bibitem[Masai(1994)]{Masai1994}   
   Masai, K. 1994, \apj, 437, 770 
\bibitem[Masui et al.(2009)]{Masui2009}
   Masui, K., Mitsuda, K., Yamasaki, N., Takei, Y., Kimura, S., Yoshino, T., \& McCammon, D. 2009, \pasj, 61, 115
\bibitem[Mitsuda et al.(2007)]{Mitsuda2007}
   Mitsuda, K., et al. 2007, \pasj, 59, S1
\bibitem[Ozawa et al.(2009)]  {Oz09}
   Ozawa, M., Koyama, K., Yamaguchi, H., Masai, K., \& Tamagawa, T. 2009, \apj,  706, L71
\bibitem[Rakowski et al.(2003)] {Rakowski2003}
   Rakowski, C. E., Ghavamian, P., \& Hughes, J. P. 2003, \apj, 590, 846
\bibitem[Rho \& Petre(1998)]{Rho1998}
    Rho, J., \& Petre, R. 1998, \apj, 503, L167
\bibitem[Sawada \& Koyama(2012)]{Sawada2012} 
   Sawada, M., \& Koyama, K. 2012, \pasj, 64, 81 
\bibitem[Sedov(1959)] {Sedov1959}  
   Sedov, L. I. 1959, Similarity and Dimensional Methods in Mechanics, New York: Academic Press, 1959
\bibitem[Serlemitsos et al.(2007)]{Se07} 
   Serlemitsos, P., et al.\ 2007, \pasj, 59, S9
\bibitem[Shimizu et al.(2013)]{Sh13} 
   Shimizu, T., Masai, K., \& Koyama, K. 2013, \pasj, 65, 69
\bibitem[Tawa et al.(2008)] {Tawa2008} 
   Tawa, N., et al. 2008, \pasj, 60, S11
\bibitem[Tsunemi et al.(1986)]{Ts86} 
   Tsunemi, H., Yamashita, K.,  Masai, K.,  Hayakawa, S., \& Koyama, K.  1986, \apj, 306, 248
\bibitem[Truelove \& McKee(1999)] {Truelove1999} 
   Truelove, J. K., \& McKee, C. F. 1999, \apjs, 120, 299.
\bibitem[Uchiyama et al.(2013)]{Uchiyama2013}
   Uchiyama, H., Nobukawa, M., Tsuru, T. G., \& Koyama, K.  2013, \pasj, 65, 19
\bibitem[White \& Long(1991)]{White1991}  
   White, R. L., \& Long, K. S. 1991, \apj, 373, 543
\bibitem[Yamaguchi et al.(2008)]{Ya08} 
   Yamaguchi, H., et al.  2008, \pasj,  60, S141
\bibitem[Yamaguchi et al.(2018)]{Ya18} 
   Yamaguchi, H., et al.  2018, \apj, 868, L35

\end{thebibliography}
\end{document}